\documentclass[prd,aps,amsmath,twocolumn,amssymb,nofootinbib,preprintnumbers]{revtex4-2}
\usepackage{graphicx,array}
\usepackage{hyperref}
\usepackage{xcolor}
\usepackage{amsmath,amssymb,slashed,latexsym}
\usepackage{bm}
\usepackage{braket}
\usepackage{placeins}
\usepackage{adjustbox}
\usepackage[caption=false]{subfig}
\usepackage{enumitem}
\usepackage{amsthm}
\usepackage{tikz}
\usepackage[compat=1.1.0]{tikz-feynman}

\DeclareMathOperator{\tr}{tr}

\DeclareRobustCommand{\Sec}[1]{Sec.~\ref{#1}}
\DeclareRobustCommand{\Secs}[2]{Secs.~\ref{#1} and \ref{#2}}

\DeclareRobustCommand{\App}[1]{App.~\ref{#1}}

\DeclareRobustCommand{\Fig}[1]{Fig.~\ref{#1}}

\DeclareRobustCommand{\Eq}[1]{Eq.~(\ref{#1})}
\DeclareRobustCommand{\Eqs}[2]{Eqs.~(\ref{#1}) and (\ref{#2})}

\DeclareRobustCommand{\Reff}[1]{Ref.~\cite{#1}}

\def\beq{\begin{equation}}
\def\eeq{\end{equation}}
\def\bea{\begin{eqnarray}}
\def\eea{\end{eqnarray}}

\DeclareMathOperator*{\argmin}{arg\,min}

\usepackage{array}
\newcolumntype{P}[1]{>{\centering\arraybackslash}p{#1}}
\newcolumntype{M}[1]{>{\centering\arraybackslash}m{#1}}

\definecolor{lightblue}{rgb}{0.1, 0.5, 1.0}
\definecolor{darkblue}{cmyk}{1,0.4,0,0.3}
\definecolor{violet}{cmyk}{0,1,0,0.2}
\hypersetup{colorlinks, bookmarksnumbered, citecolor=darkblue, linkcolor=darkblue, pdfstartview=FitH, urlcolor=darkblue, linktocpage}

\begin{document}

\preprint{MIT-CTP/5788}

\title{Flavor Patterns of Fundamental Particles from Quantum Entanglement?}

\author{Jesse Thaler}\email{jthaler@mit.edu}
\affiliation{Center for Theoretical Physics, Massachusetts Institute of Technology, Cambridge, MA 02139, USA}
\author{Sokratis Trifinopoulos}\email{trifinos@mit.edu}
\affiliation{Center for Theoretical Physics, Massachusetts Institute of Technology, Cambridge, MA 02139, USA}

\begin{abstract}

The Cabibbo–Kobayashi–Maskawa (CKM) matrix, which controls flavor mixing between the three generations of quark fermions, is a key input to the Standard Model of particle physics.
In this paper, we identify a surprising connection between quantum entanglement and the degree of quark mixing.
Focusing on a specific limit of $2 \to 2$ quark scattering mediated by electroweak bosons, we find that the quantum entanglement generated by scattering is minimized when the CKM matrix is almost (but not exactly) diagonal, in qualitative agreement with observation.
With the discovery of neutrino masses and mixings, additional angles are needed to parametrize the Pontecorvo–Maki–Nakagawa–Sakata (PMNS) matrix in the lepton sector.
Applying the same logic, we find that quantum entanglement is minimized when the PMNS matrix features two large angles and a smaller one, again in qualitative agreement with observation, plus a hint for suppressed CP violation.
We speculate on the (unlikely but tantalizing) possibility that minimization of quantum entanglement might be a fundamental principle that determines particle physics input parameters. 

\end{abstract}

\maketitle

\tableofcontents

\section{Introduction}
\label{sec:intro}

The Standard Model (SM) of particle physics---which provides our most precise understanding of microscopic physics in our Universe---has 19 input parameters related to the masses, mixings, and couplings between quarks, leptons, gauge bosons, and the Higgs boson.
A grand challenge in fundamental physics is to understand the origins of, and relations between, these 19 parameters.
The most notable example of possible parameter reduction appears in the gauge sector, where the strong, electroweak, and hypercharge gauge couplings could have a common origin in grand unified theories~\cite{Georgi:1974sy,Pati:1974yy,Buras:1977yy}.
Just as the dizzying spectrum of strongly-interacting hadrons was explained in terms of constituent quarks and gluons in quantum chromodynamics (QCD)~\cite{Gell-Mann:1961omu,Gell-Mann:1964ewy,Han:1965pf,Fritzsch:1973pi,Gross:1973id,Politzer:1973fx}, the puzzle of SM parameters could be resolved by finding new high-energy structures.

An alternative possibility is that the explanation for SM parameters could arise outside of the traditional framework of quantum field theory.
Inspired by the landscape of string theory vacua, an increasingly mainstream view is that some SM parameters might be determined by environmental selection (a.k.a.~the anthropic principle)~\cite{Weinberg:1987dv}.
Not all SM parameters can be explained in this way, however, and the fact that the SM features three generations of fermions is particularly perplexing given that one generation would naively suffice for supporting a habitable Universe.
Indeed, the curious patterns of fermion masses and mixing in the quark and lepton sectors motivates detailed investigations into flavor physics~\cite{Altmannshofer:2022aml}.
\begin{figure}
    \centering
    \includegraphics[width=0.8\linewidth]{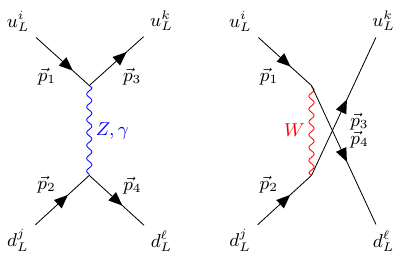}
     \caption{
    Two competing SM Feynman diagrams with respect to flavor entanglement in $u_L d_L \to u_L d_L$ elastic scattering at tree level.
    Left: Flavor preserving via $t$-channel exchange of a $Z$ boson or a photon.  Right: Flavor changing via $u$-channel exchange of a $W$ boson.}
    \label{fig:diagrams}
\end{figure}
In this paper, we encounter the surprising possibility that quantum entanglement could provide an explanation for the flavor structure of SM fermions.
Specifically, we calculate the amount of quantum entanglement generated in electroweak $2 \to 2$ scattering processes between left-handed quarks of various flavors, as shown in \Fig{fig:diagrams}:
\begin{equation}
\label{eq:ud_channel}
    u_{Li} d_{Lj} \to u_{Lk} d_{L\ell}\, .
\end{equation}
By minimizing a specific measure of entanglement at an energy scale relevant to the particles of interest, we find that the three quark mixing angles in the Cabibbo–Kobayashi–Maskawa (CKM) matrix exhibit a \emph{hierarchical} flavor structure:
\begin{equation}
45^\circ \gg \theta_{{\rm CKM},12} > \theta_{{\rm CKM},13} \approx \theta_{{\rm CKM},23} \approx 0\, .
\end{equation}
Extending this analysis to leptons, at least 7 parameters are required beyond the 19 of the SM to explain neutrino masses, with lepton flavor mixing encoded in the Pontecorvo–Maki–Nakagawa–Sakata (PMNS) matrix.
Minimizing the analogous entanglement in lepton scattering, $\nu_{L} \ell_{L} \to \nu_{L} \ell_{L}$,
we find that the three lepton mixing angles are consistent with the flavor structure:
\begin{equation}
45^\circ > \theta_{{\rm PMNS}, 12} \sim \theta_{{\rm PMNS},23} > \theta_{{\rm PMNS},13} \gg 0\, .
\end{equation}
We emphasize that we are using a common entanglement principle, but quarks and leptons nevertheless exhibit different flavor structures due to their different electroweak charges.

Entanglement is a core phenomenon in quantum mechanics, where measurement outcomes are correlated beyond classical expectations.
In particle physics, entanglement is so ubiquitous that we often take it for granted, but every neutral pion decay to two photons ($\pi_0 \to \gamma \gamma$) is effectively a mini Einstein--Podolsky--Rosen experiment.
In the context of SM scattering processes, though, the study and quantification of entanglement in its own right has only begun relatively recently~\cite{Balasubramanian:2011wt, Seki:2014cgq, Peschanski:2016hgk,Grignani:2016igg,Kharzeev:2017qzs,Fan:2017hcd,Fan:2017mth,Beane:2018oxh,Rigobello:2021fxw,Low:2021ufv,Liu:2022grf,Fedida:2022izl,Cheung:2023hkq,Carena:2023vjc,Aoude:2024xpx,Low:2024mrk,Low:2024hvn}.
In terms of predicting particle properties from entanglement, the two papers we are aware of are: i) \Reff{Cervera-Lierta:2017tdt} which showed that maximizing helicity entanglement yields a reasonable prediction for the Weinberg angle $\theta_W$ that controls the mixing between electroweak bosons, and ii) \Reff{Quinta:2022sgq} which showed that minimizing entanglement in neutrino oscillations yields a prediction for the CP-violating phase of the PMNS matrix.

Our study of entanglement for flavor physics was motivated by recent work on the interplay between entanglement in particle scattering and the emergence of global symmetries at low energies.
 Refs.~\cite{Beane:2018oxh,Low:2021ufv,Liu:2022grf} showed that the suppression of spin-entanglement in non-relativistic baryon scattering leads to enlarged symmetries, such as Wigner's $SU(4)_{sm}$ spin-flavor symmetry~\cite{Wigner1995} for two flavors, $SU(16)$ for three, and the Schr\"odinger conformal invariance~\cite{Mehen:1999nd,Nishida:2007pj}.
\Reff{Carena:2023vjc} examined the entanglement produced by scattering in a generic two-Higgs doublet model, finding that vanishing entanglement leads to the maximal symmetry $SO(8)$.%
\footnote{\Reff{Kowalska:2024kbs,Chang:2024wrx} recently highlighted the importance of analyzing the entanglement properties of all different scattering channels between the external particles.}
Based on these previous studies, we expected that suppressing entanglement for scattering involving flavor would again lead to enhanced symmetries, for example an enhanced $SU(3)$ flavor symmetry rotating all three quark generations.

Instead, the SM had something different in store for us, as interference between the diagrams in \Fig{fig:diagrams} leads to non-trivial flavor symmetry breaking effects when trying to minimize entanglement.
Our measure of entanglement depends on the scattering kinematics, and we use the scattering angle $\Theta = 90^\circ$ as a baseline value and set the scattering energy $\sqrt{s}$ to be between the heaviest fermion in the spectrum and the electroweak scale.
With these choices, we find that the CKM matrix that minimizes entanglement is approximately diagonal albeit with a non-zero Cabibbo angle $\theta_C \equiv \theta_{{\rm CKM},12}$, consistent with models featuring an approximate $U(2)$ flavor symmetry in the quark sector~\cite{Barbieri:1995uv,Barbieri:1996ww,Barbieri:2011ci}.
By contrast, the PMNS matrix that minimizes entanglement is far from diagonal, with all entries of comparable size, reminiscent of the \emph{anarchic} flavor hypothesis in the lepton sector~\cite{Hall:1999sn,Haba:2000be}.
These results could simply be a coincidence, of course, or they could be a harbinger for a new fundamental principle that determines flavor mixing parameters.

Regardless of whether entanglement minimization is a principle of nature, the appearance of \emph{any} prediction for a hierarchical CKM matrix that simultaneously predicts an anarchic PMNS matrix is cause for celebration (or alarm, depending on your perspective).
The quark and lepton flavor sectors of the SM have been notoriously difficult to decipher.
In pursuit of an underlying organizing principle, spurion analysis has been employed, often motivated by numerical relations among the fermion masses and CKM elements~\cite{Gatto:1968ss,Koide:1982ax,Xing:2002sw,Grossman:2020qrp}. 
Traditionally, these investigations have led to the proposal of flavor symmetries at high energies that appear broken in the low-energy limit~\cite{Froggatt:1978nt,Chivukula:1987py,Barbieri:1995uv,Barbieri:1996ww,DAmbrosio:2002vsn,He:2006dk,Alonso:2013nca}.
For example, in Froggatt–Nielsen (FN) models~\cite{Froggatt:1978nt,Leurer:1992wg,Ibanez:1994ig}, the structure of the CKM matrix is determined by $U(1)$ symmetries under which different fermion flavors have different charges.
Such models can help explain why the Cabibbo angle $\theta_C$ is small but non-zero, since it arises from a ratio of mass scales. 
This kind of spurion analysis, however, is not well matched to the structure of the PMNS matrix.
To our knowledge, this is the first time the differing CKM and PMNS structures have arisen from a common mechanism that does not invoke symmetries or symmetry breaking.

The remainder of this paper is structured as follows.
In section \ref{sec:setup}, we review the concept of entangling power from quantum information theory and explain its relevance to electroweak scattering of quarks.
In section \ref{sec:minimization}, we minimize the flavor entanglement induced by electroweak interactions with respect to the angles in the CKM and PMNS matrices.
We discuss future extensions and fever dreams in section~\ref{sec:conclusions}.

\section{Setup}
\label{sec:setup}

\subsection{Flavor Hilbert Space}
\label{sec:Hilbert_flavor}

Let us first consider the Hilbert space corresponding to quark flavor.
For up quarks, we consider a $G$-dimensional Hilbert space $H_u$, where $G$ is the number of generations, and similarly for the down quarks $H_d$.
In the full three-generation scheme ($G = 3$), the states are qutrits with the following basis elements:
\begin{align} \label{eq:Hilbert_ud}
    H_u: \quad & \ket{1}_u\, , \quad \ket{2}_u, \quad \ket{3}_u\, ,\\
    H_d: \quad & \ket{1}_d\, , \quad \ket{2}_d, \quad \ket{3}_d\, .
\end{align}
These correspond, respectively, to the quark flavors up, charm, top, down, strange, and bottom. 
The anti-particles belong to the complex conjugate Hilbert spaces $H_{\bar{u}}$ and $H_{\bar{d}}$, which are, however, isomorphic to $H_{u}$ and $H_{d}$. 
Similarly, we can define Hilbert spaces for the leptons $H_\ell$ and neutrinos $H_\nu$.

Throughout our discussion, whenever we refer to a state of a given ``flavor,'' we really mean a mass eigenstate.
This is particularly relevant for the discussion of neutrinos, where the flavor eigenstates (i.e.\ electron-, muon-, and tau-type neutrinos) are substantially different from the mass eigenstates (i.e.~$\nu_1$, $\nu_2$, and $\nu_3$).
Since we most often refer to states by their index $i = 1, \ldots, G$, we hope that no confusions arise. 

Considering now the product Hilbert space,
\begin{equation}
\label{eq:full_hilbert}
    H_f=H_u \otimes H_d\, ,
\end{equation}
a generic state can be written as
\begin{equation}
\label{eq:generic_state}
    \ket{\alpha} = \sum_{i,j = 1}^G \alpha_{ij} \ket{ij}_{ud}\, , \qquad \ket{ij}_{ud} \equiv \ket{i}_u \otimes \ket{j}_d\, ,
\end{equation}
where $\alpha$ is a $G \times G$ matrix with $\tr(\alpha^\dagger \alpha) = 1$ to ensure normalization.
If $\alpha$ is a rank-1 matrix, then $\ket{\alpha}$ takes the form of an unentangled product state.

Viewing $\ket{\alpha}$ as a generic bipartite state, we can quantify the entanglement between the up-type and down-type qubits in various ways.
A convenient measure of entanglement is the linear entropy~\cite{Zanardi:2000zz}.%
\footnote{This quantity measures the purity of the reduced density matrix $\rho_R$ and it is in general not a formal entanglement measure (see \Reff{Plenio:2007zz} for a comprehensive review). For pure states as considered here, however, it can be used to characterize entanglement, since it is the linear limit of the von Neumann entropy, which is the unique entanglement measure for pure bipartite systems~\cite{Bennett:1995tk}.
For the two-flavor study in \Sec{sec:2FS}, we checked that entanglement minimization yields the same result for linear entropy and von Neumann entropy.}
For a bipartite system with density matrix $\rho$, the linear entropy is defined as
\begin{equation} \label{eq:lin_entropy}
    E(\rho) \equiv \frac{G}{G-1}\left| 1 - \tr{\rho_{R}^2} \right|\, ,
\end{equation}
where $\rho_{R}$ is the reduced density matrix across the partition.
This number ranges from 0 (for no entanglement) to 1 (maximal entanglement).
For convenience when working with pure states, we will often use the notation $E(\ket{\alpha})$.
For the special cases of $G = 2$ and $G=3$:
\begin{align}
E(\ket{\alpha})_{G = 2} &= 4 \lambda_1 \lambda_2\, , \\
E(\ket{\alpha})_{G = 3} &= 3 \left(\lambda_1 \lambda_2 + \lambda_2 \lambda_3 + \lambda_3 \lambda_1 \right)\, ,
\end{align}
where $\lambda_i$ are the eigenvalues of $\alpha^\dagger \alpha$ that satisfy the trace condition $\sum_i \lambda_i = 1$.

\subsection{Entangling Power}
\label{sec:entangling_power}

Starting from the unentangled product state, we can ask how much entanglement is generated by an operator $\mathcal{S}_f$, with a notation inspired by (but not identical to) the scattering operator.
The \emph{entangling power} of $\mathcal{S}_f$ quantifies how entangled the final state is after applying $\mathcal{S}_f$ to the initial state~\cite{Zanardi:2000zz}.
Using the linear entropy in \Eq{eq:lin_entropy} as a measure of state entanglement, the entangling power of $\mathcal{S}_f$ is defined as
\begin{equation}
\label{eq:entangling_power_def}
\mathcal{E}(\mathcal{S}_f) \equiv \overline{ E\big(\mathcal{S}_f\ket{i}_u \otimes \ket{j}_d\big)}\, ,
\end{equation}
where the bar denotes the average over a chosen set of initial product states.

The focus of this paper is on $2\to 2$ left-handed fermion (or, after crossing, right-handed anti-fermion) scattering:
\begin{equation}
    \label{eq:flavor_kinematics}
    f_L^i(p_1) f_L^j(p_2) \to f_L^k(p_3) f_L^\ell(p_4)\, ,
\end{equation}
where $p$ are the corresponding four-momenta of the particles.
When we include fermion masses, the subscript ``$L$'' (``$R$'') will refer to negative (positive) helicity particles.
With fixed helicities, the relevant amplitudes depend on the kinematics of the external particles: the center-of-mass energy $\sqrt{s}$ and the scattering angle $\Theta$ between the incoming and outgoing particles, as discussed further in \Sec{sec:channels_kinematics}.%
\footnote{Because we are dealing with initial-state left-handed fermions whose spin is anti-aligned with its momentum, there is no dependence on the azimuthal angle around the beam direction.}

In general, $2 \to 2$ fermion scattering will generate entanglement just from momentum and helicity correlations, but our goal is to isolate the flavor component.
We therefore consider a situation where we \emph{prepare} the initial-state particles with certain momenta, helicities, and flavors, and \emph{measure} the helicities and kinematics of the final-state particles (but not the flavor).
Concretely, we project down to final-state particles whose spin is anti-aligned with their momentum (negative helicity) and anti-particles with aligned spin and momentum (positive helicity), and consider final-state particles that appear at a given scattering angle $\Theta$, without making any flavor projections.%
\footnote{This implies the ability to measure helicity without measuring mass, which is possible for a thought experiment, though challenging for a real one.}
The formal description of this procedure can be found in \App{app:scattering_details}.

We denote the combined action of scattering and measurement by $\mathcal S_f$, which should not be confused with the unitary scattering operator $\mathcal S$ in Fock space.
In our case, the operator $\mathcal S_f$ is not unitary---both because of the restriction to two-particle final states and because of the projections above---and it acts on the flavor Hilbert space $H_f$.
As explained in \App{app:scattering_details}, one can define $\mathcal S_f$ such that it preserves the norm for specific scattering channels, though orthogonal initial states will in general map to non-orthogonal final states.%

Even though $\mathcal S_f$ is not unitary, we can still use the entangling power in \Eq{eq:lin_entropy} to reliably characterize entanglement of the final state.
The reason is that we are focusing on \emph{elastic} scattering, i.e.\ transitions between initial and final states of the same SM quantum numbers (up to flavor).
Therefore, we can consider an idealized measurement that discards final states that do not have the desired kinematics but nevertheless leaves flavor quantum coherence intact, such that no classical admixture is generated by $\mathcal S_f$.
In this way, scattering maps states within the same flavor Hilbert space:
\begin{equation}\label{eq:scriptS_def}
   H_{f} \xrightarrow{\mathcal S_f} H_{f}\, .
\end{equation}

Next, we have to choose the states to average over in \Eq{eq:entangling_power_def}.
In the absence of fermion masses, the CKM and PMNS matrices can be diagonalized by doing a chiral rotation.
This, in turn, means that \Eq{eq:entangling_power_def} will have no sensitivity to flavor mixing angles in the massless limit, unless the average depends on a preferred choice of flavor frame.
We therefore choose to average over product states of definite fermion generation:
\begin{equation}
\label{eq:flavor_averaging}
\mathcal{E}_{ud}(\mathcal{S}_f) = \frac{1}{G^2}\sum_{i,j = 1}^G E\big(\mathcal{S}_f \ket{ij}_{ud} \big)\, .
\end{equation}
This quantity depends on SM parameters through $\mathcal{S}_f$, and it also depends on the choice of particular elastic channel and scattering kinematics, as we now discuss.

\subsection{Scattering Channels and Kinematics}
\label{sec:channels_kinematics}

The $u d \to u d$ diagrams in \Fig{fig:diagrams} correspond to the lowest-order flavor-changing electroweak scattering processes, but we can also consider its crossed version:
\begin{align}
    u_{Li} \bar{d}_{Rj} \to u_{Lk} \bar{d}_{R\ell}\, .
\end{align}
Note that we do not consider the $u \bar{u} \to \bar{d} d$ channel since it is inelastic.
There are also conjugate processes with $\bar{u}_{Ri}$ as the first particle, but they do not provide additional information because of CPT symmetry.

Because entangling power is not crossing symmetric, we have to decide which channel(s) to study.
We choose to analyze the minimum electroweak entangling power:
\begin{equation}
\label{eq:min_entanglment_power}
\mathcal{E}_{\rm min}(\mathcal{S}_f) \equiv \min \big(\mathcal{E}_{u d},\mathcal{E}_{u \bar{d}}\big)\, ,
\end{equation}
where the subscripts correspond to the initial state of the scattering channel.

We then have choose the kinematics of the scattering.
We focus on perpendicular scattering with
\begin{equation} 
\label{eq:t_u_choice}
\Theta = \frac{\pi}{2}\, ,
\end{equation}
and indicate this choice with a $\perp$ superscript:
\begin{equation}
\label{eq:perp_entangle_power}
\mathcal{E}_{\rm min}^\perp(\mathcal{S}_f^\perp) \equiv \mathcal{E}_{\rm min}(\mathcal{S}_f) \Big\vert_{\Theta = \frac{\pi}{2}}\, .
\end{equation}
We refer to this quantity as the \emph{perpendicular entangling power}, which has the nice feature of being invariant to spatial reversal of the two incident beams.%
\footnote{An alternative way to respect spatial inversion is including $\mathcal{E}_{d u}$ and $\mathcal{E}_{\bar{d} u}$ terms in the minimization of \Eq{eq:min_entanglment_power}.
We stress that we do not impose invariance with respect to spatial inversion as a symmetry at the Lagrangian level (which is obviously violated by electroweak interactions) or amplitude level (which also does not hold for distinguishable particles).
Instead, we are requiring that the result of the minimization processes over the possible range of SM couplings is invariant under this transformation, which constrains the ``ensemble'' of different versions of the SM in the space of possible couplings.}
For massless particles, perpendicular scattering corresponds to the Mandelstam variables
$t=u=-\frac{s}{2}$,  which again emphasizes that entangling power is not crossing symmetric.

In the analysis below, we explore different choices for the center-of-mass collision energy $\sqrt{s}$, finding the most interesting behavior when $\sqrt{s}$ is between the heaviest participating fermion and the electroweak scale:
\begin{equation}
\label{eq:sqrt_s_choice}
m_Z \lesssim \sqrt{s} \lesssim m_t~ \text{(quarks)\,,} \quad m_{\tau} \lesssim \sqrt{s} \lesssim m_Z~ \text{(leptons)}~.
\end{equation}
Even though our prescription yields the same qualitative results across these energy ranges, when presenting the final results in \Secs{sec:3FS_CKM}{sec:3FS_PMNS}, we choose the value of $\sqrt{s}$ that minimizes $\mathcal{E}^\perp_{\rm min}$ within those intervals.

\subsection{Alternative Choices}

Ultimately, we will identify CKM- and PMNS-like structures by minimizing \Eq{eq:perp_entangle_power} with respect to the flavor mixing parameters.
Given the numerous choices we made---particularly the flavor averaging in \Eq{eq:flavor_averaging}, the scattering channel minimization in \Eq{eq:min_entanglment_power}, the scattering kinematics in \Eqs{eq:t_u_choice}{eq:sqrt_s_choice}, and the choice to minimize (as opposed to maximize) entanglement---one can reasonably ask about a ``look elsewhere'' effect.
After all, \Reff{Cervera-Lierta:2017tdt} focused on \emph{maximizing} helicity entanglement, whereas we are focusing on \emph{minimizing} flavor entanglement.

For completeness, we explored alternative choices for the entanglement measure, by replacing the default choices above with various combinations of minimization, maximization, and averaging over the relevant quantities.
In the two-generation context, we found that alternative choices tended to yield values of the Cabbibo angle of either $0^\circ$ or $45^\circ$, qualitatively different from the SM value of $13^\circ$.
Of the approaches we tested, only the specific choices presented in this section allow for a non-trivial interplay between the processes in \Fig{fig:diagrams}.

\section{Entangling Power of Electroweak Interactions}
\label{sec:minimization}

For the following analysis, we study the perpendicular entangling power $\mathcal{E}^\perp_{\rm min}$ as a function of the center-of-mass collision energy $\sqrt{s}$ and the flavor mixing angles.
We use the standard parametrization of the CKM and PMNS matrices, as reviewed in \App{app:standard_parametrization}.
Except where noted, gauge couplings, gauge boson masses, and fermion masses are set to their SM values.

We start with the case of two quark generations to gain intuition, and then present three-generation results.
The relevant scattering amplitudes are reviewed in \App{app:amplitudes}.

\begin{figure*}
    \centering
    \subfloat[]{
        \includegraphics[width=0.45\textwidth]{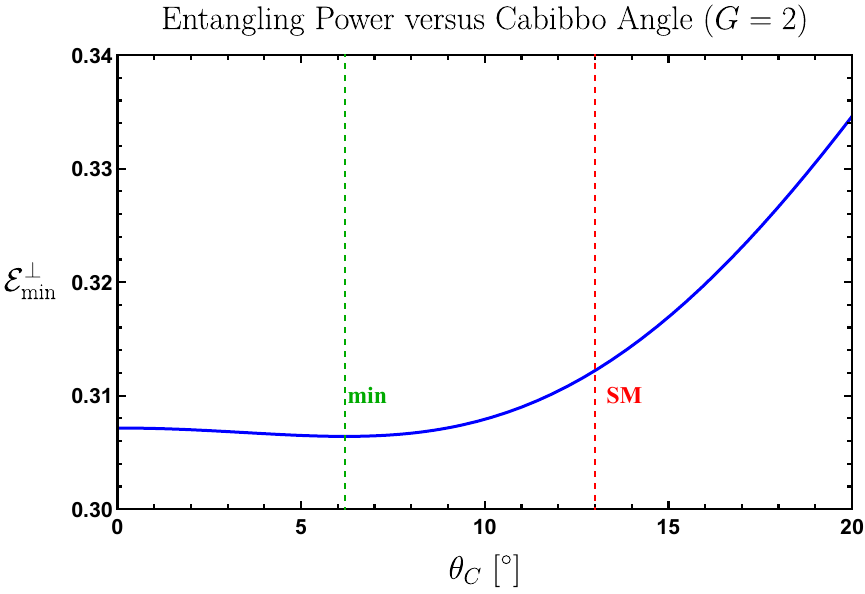} 
        \label{fig:min}
    }
    $\qquad$
    \subfloat[]{
     \includegraphics[width=0.45\textwidth]{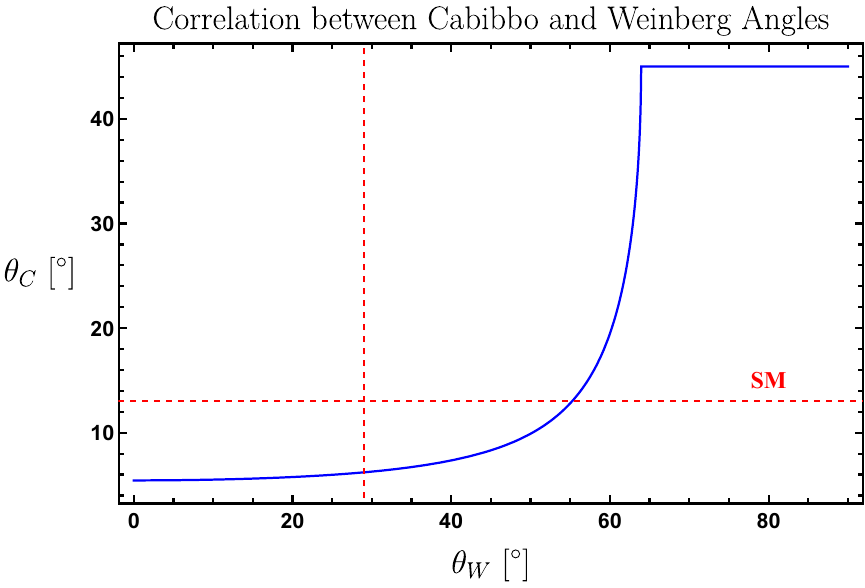}
        \label{fig:thetaC_thetaW}
    }
    \caption{
    (a)  The orthogonal entangling power (which is dominated by the $ud\to ud$ channel) as a function of the Cabibbo angle.
    The position of the entangling power minimum is indicated by the dashed green line while the SM value by the red.
    (b)   Correlation between the Cabibbo and the Weinberg angles at the entangling power minimum.
    The SM values are indicated by the red dashed lines.
    }
\end{figure*}

\subsection{The Cabibbo Angle}
\label{sec:2FS} 

In the two-flavor limit for quarks, we can set $\theta_{13}=\theta_{23}=0$ in \Eq{eq:SP}, such that flavor-changing effects in the quark sector are governed solely by the Cabibbo angle $\theta_{12}=\theta_C \in [0,\pi/4]$.%
\footnote{In principle, we could perform a similar exercise for the PMNS matrix, though without an approximate $U(2)$ flavor symmetry, retaining only one angle is not a useful limit.}
We want to understand whether the Cabibbo angle can be recovered by minimizing the perpendicular entangling
power:
\begin{equation}
\label{eq:min_ent_strength}
    \theta_C^{\rm min} = \argmin_{\rm ch,\theta_C} \mathcal E_{\rm ch}^{\rm \perp} [\theta_C ]\, . 
\end{equation}

Empirically, we find that the minimum in \Eq{eq:min_ent_strength} is achieved for the $ud \to ud$ channel, which is also the channel that sets the three-generation $\mathcal{E}^\perp_{\rm min}$ at the nominal choice of $\sqrt{s} \lesssim m_t$, as we study in \Sec{sec:3FS_CKM} below.
At leading order, $ud \to ud$ is induced by $Z/\gamma$ boson exchange in the $t$-channel, and $W$ boson exchange in the $u$-channel, as shown in \Fig{fig:diagrams}.
This process depends on the Weinberg angle $\theta_W$, which controls the mixing of the $SU(2)_L$ and $U(1)_Y$ gauge fields to produce the $Z$ boson.
This mixing also defines the ratio of the $W$ and $Z$ boson masses, namely:
\begin{equation}
\cos\theta_W = \frac{m_W}{m_Z}\, .
\end{equation}
We will find a fascinating dependence of $\theta_C^{\rm min}$ on $\theta_W$ in \Fig{fig:thetaC_thetaW} below.

Following \App{app:amplitudes} and taking the massless quark limit,%
\footnote{ In the massless limit, the Cabibbo angle can be removed by a field redefinition.  Because of the specific flavor averaging procedure in \Eq{eq:flavor_averaging}, though, $\theta_C$ remains physical in our analysis.}
we can write the scattering amplitude as: 
\begin{align}\label{eq:ampl}
    i \mathcal M_{k\ell ij} &= -i \left( P_{k\ell ij}^{t(Z,\gamma)} + P_{k\ell ij}^{u(W)} \right) 2s \, , 
\end{align}
and the individual diagrams yield:
\begin{align}
\label{eq:P_Z_gamma}
&P_{k\ell ij}^{t(Z,\gamma)} = g^2 \delta_{ik}\delta_{j\ell} \left(\frac{Y^uY^d}{\cos^2\theta_W}\frac{ 1}{t-m_Z^2} + \frac{\sin^2\theta_W \, Q^uQ^d}{t}\right),
\\
\label{eq:P_W}
&P_{k\ell ij}^{u(W)} = \frac{g^2 V_{i\ell}^{\ast}V_{kj}}{2}  \frac{1}{u-m_W^2}\, .
\end{align}
Here, the $SU(2)_L$ gauge coupling is $g$, the weak-hypercharge is defined as
\begin{equation} \label{eq:weak_hypercharge}
   Y^f = T_3^f - Q^f \sin\theta_W^2 \, ,
\end{equation}
and the quark electric charges and weak isospins are
\begin{equation} \label{eq:quark_charges}
Q^u = \frac{2}{3}\,, \quad Q^d = -\frac{1}{3}\,, \quad T_3^u = \frac{1}{2}\,, \quad T_3^d = -\frac{1}{2}\, .
\end{equation}
Empirically, we find only mild dependence on the gauge boson masses, so to simplify the equations below, we set $m_W = m_Z = 0$, keeping $\cos\theta_W$ fixed.

To convert this amplitude to the perpendicular entangling power, we fix the kinematics to $t=u=-\frac{s}{2}$ and normalize the final-state wavefunctions in each $\ket{ij}_{ud}$ channel separately.
In the $4$-dimensional $H_f$ space for two generations, following the definition in \Eq{eq:Sf_def}:
\begin{equation} \label{eq:Sf_2FS}
\mathcal{S}_f^\perp = 
\begin{pmatrix}
\frac{2y + c_{\theta}^2}{\mathcal{N}_+} & \frac{c_{\theta} s_{\theta}}{\mathcal{N}_-} & -\frac{c_{\theta} s_{\theta}}{\mathcal{N}_-} & -\frac{s_{\theta}^2}{\mathcal{N}_+} \\
\frac{c_{\theta} s_{\theta}}{\mathcal{N}_+} & \frac{2y+s_{\theta}^2}{\mathcal{N}_-} & \frac{c_{\theta}^2}{\mathcal{N}_-} & \frac{c_{\theta} s_{\theta}}{\mathcal{N}_+} \\
-\frac{c_{\theta} s_{\theta}}{\mathcal{N}_+} & \frac{c_{\theta}^2}{\mathcal{N}_-} & \frac{2y+s_{\theta}^2}{\mathcal{N}_-} & -\frac{c_{\theta} s_{\theta}}{\mathcal{N}_+} \\
-\frac{s_{\theta}^2}{\mathcal{N}_+} & \frac{c_{\theta} s_{\theta}}{\mathcal{N}_-} & -\frac{c_{\theta} s_{\theta}}{\mathcal{N}_-} & \frac{2y + c_{\theta}^2}{\mathcal{N}_+}
\end{pmatrix}\, ,
\end{equation}
where $s_{\theta} = \sin\theta_C$, $c_{\theta} = \cos\theta_C$, and we define: 
\begin{align}
\mathcal{N}_\pm &= \sqrt{1 + 2 y + 4 y^2 \pm 2 y \cos 2 \theta_C }\, ,\\
    y &= \frac{Y^uY^d}{\cos^2\theta_W} + \sin^2\theta_W Q^u Q^d\, .
\end{align}
Note that the operator in \Eq{eq:Sf_2FS} takes an initial state of definite flavor $\ket{ij}_{ud}$ to a pure final state, i.e.\ whose density matrix $\rho$ satisfies $\rm Tr (\rho^2) = 1$. As a result, non-vanishing linear entropy implies the generation of entanglement.

For each of the four flavor initial states, we use \Eq{eq:lin_entropy} and obtain:
\begin{align}  \label{eq:entropy_same_G}
E\big(\mathcal{S}_f^\perp \ket{11}_{ud} \big) &= E\big(\mathcal{S}_f^\perp \ket{22}_{ud} \big) = 16 \frac{y^2\sin^4\theta_C}{\mathcal{N}_+^4}\, ,  \\ \label{eq:entropy_different_G}
E\big(\mathcal{S}_f^\perp \ket{12}_{ud} \big) &= E\big(\mathcal{S}_f^\perp \ket{21}_{ud} \big) = 16 \frac{y^2\cos^4\theta_C}{\mathcal{N}_-^4}\, .
\end{align}
Finally, performing the flavor averaging in \Eq{eq:flavor_averaging} with $G = 2$:
\begin{align}
\label{eq:entangling_power_2FS_ud}
    \mathcal{E}_{ud}^{\perp} &= 8 y^2 \Big[\frac{\cos^4 \theta_C}{(1 + 2 y + 4 y^2 - 2 y \cos 2 \theta_C )^2} \notag \\
    &\qquad + \frac{\sin^4 \theta_C}{(1 + 2 y + 4 y^2 + 2 y \cos 2 \theta_C )^2}\Big]\, .
\end{align}
This expression is plotted as a function of $\theta_C$ in \Fig{fig:min}, where we see that the minimum is at $\theta_C^{\rm min} = 6^{\circ}$.
This is close to the experimentally established SM value of $\theta_C^{\rm exp} \approx 13^{\circ}$.%
\footnote{We happened to notice that in the limit where we neglect photon exchange, the exact value $\theta_C^{\rm min} = 13^{\circ}$ is recovered.
However, we do not have a good reason on quantum field theoretic grounds to neglect the photon contribution.
Because of the shallow entanglement minimum in \Fig{fig:min}, a 10\% increase in the charged-current process over the neutral-current one would be enough to accomplish this shift, which is roughly of the expected size for higher-order corrections.}

We can understand how the non-trivial flavor-mixing angle arises from the minimization by observing the following relative imbalance: the $W$-induced charged-current process generates both flavor-changing and flavor-conserving effects, while the $\gamma/Z$-induced neutral-current process only flavor-conserving effects. 
In fact, according to \Eq{eq:P_W} the CKM factor that appears in the charged-current term takes different values depending on whether the flavor of each quark has changed or not:
\begin{equation}
V_{i\ell}^{\ast}V_{kj} = 
\begin{cases}
\cos^2 \theta_C & i=\ell \text{ and } j=k\, , \\
\sin^2 \theta_C & i \neq \ell \text{ and } j \neq k\, 
, \\
\sin \theta_C \cos \theta_C & i \neq \ell \text{ xor } j \neq k\, .
\end{cases}
\end{equation}
When taking both processes into account, in case the initial quarks are of the same generation the flavor-changing effects enhance flavor entanglement, while in case the initial quarks are of different generations they suppress it (see \Eqs{eq:entropy_same_G}{eq:entropy_different_G}).

Further inspecting \Eq{eq:entangling_power_2FS_ud}, we see that the Weinberg angle $\theta_W$ is the only other SM coupling that affects the value of the Cabibbo angle $\theta_C^{\rm min}$ at the entangling power minimum.
We study this dependence of $\theta_C^{\rm min}$ on $\theta_W$ in \Fig{fig:thetaC_thetaW}.
As $\sin\theta_W \to 0$, the contribution from the flavor-changing $W$-boson exchange becomes larger, and thus minimizing entanglement requires a smaller (though non-vanishing) value for $\theta_C^{\rm min}$.
The minimum is then a compromise between the flavor-changing effects of the $W$ boson that generate entanglement and those that suppress it, as explained above.
In the opposite limit $\sin\theta_W \to 1$, the $Z$-boson exchange dominates and \Eq{eq:entangling_power_2FS_ud} simplifies to $\mathcal{E}_{ud}^{\perp}  = (\cos^4 \theta_C + \sin^4 \theta_C) /(2y^2)$. Therefore $\theta_C = \pi/4$ becomes the preferred minimum. 

\subsection{Towards the Full CKM Matrix}
\label{sec:3FS_CKM}

\begin{figure*}
\centering
\subfloat[]{
    \includegraphics[width=0.45\textwidth]{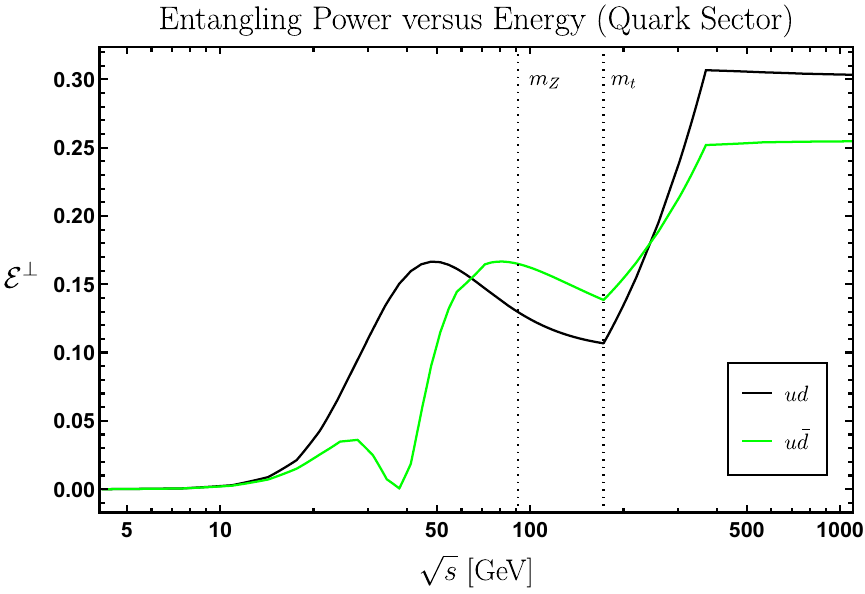} 
    \label{fig:EP_vs_energy_quarks}
}
$\qquad$
\subfloat[]{
 \includegraphics[width=0.45\textwidth]{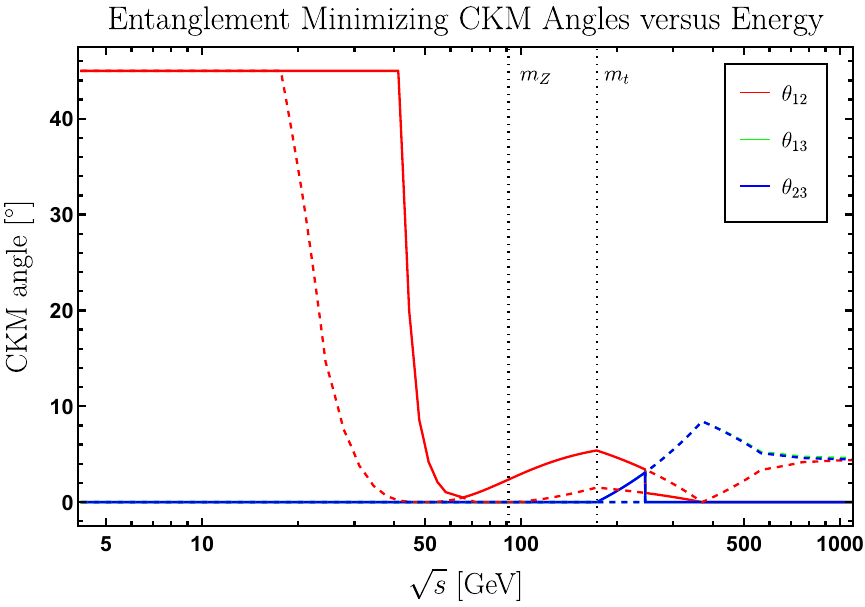}
    \label{fig:angles_vs_energy_quarks}
}
\caption{
Electroweak entanglement in the quark sector as a function of the scattering center-of-mass energy $\sqrt{s}$.
(a)
The perpendicular entangling power after minimizing over the CKM angles, considering the two scattering channels $ud \to ud$ and $u\bar{d} \to u\bar{d}$ separately.
(b)
The CKM angles at the entangling power minimum, with solid lines corresponding to the channel that minimizes $\mathcal{E}^{\perp}$ at each $\sqrt{s}$ and dashed lines to the other channel for reference.
The CP-violating phase $\delta_{\rm CKM}$ has no impact on the minimization.
}
\end{figure*}

Extending the above analysis for three generations of quarks is straightforward.
A key difference, though, is that we need to keep track of fermion mass dependence, such that only quarks that can be kinematically accessed at a given center-of-mass energy $\sqrt{s}$ contribute to the scattering.%
\footnote{Since $\mathcal{S}_f$ is already a non-unitary operator, there are no additional complications from discarding kinematically forbidden channels.
To avoid discontinuous behavior, though, we use the $G = 3$ normalization of \Eq{eq:flavor_averaging} even when only a subset of flavors participate in the scattering.
}
We stress that the energy dependence of the entangling power should not be confused with renormalization group evolution, which is practically negligible in the studied energy ranges.

Considering the $ud \to ud$ and $u\bar{d} \to u\bar{d}$ channels separately, we present the perpendicular entangling power $\mathcal{E}^\perp$ as a function of $\sqrt{s}$ in \Fig{fig:EP_vs_energy_quarks}.
Here, we have minimized over the CKM angles for each channel separately.
The two channels alternate as to which sets the minimum in \Eq{eq:min_entanglment_power}.
As anticipated in \Sec{sec:2FS}, the $ud \to ud$ channel sets the minimum in the range $m_Z \lesssim \sqrt{s} \lesssim m_t$.

In \Fig{fig:angles_vs_energy_quarks}, we present the CKM angles that minimize $\mathcal{E}^\perp_{\rm min}$ in solid lines, with the angles that minimize the other channel shown in dashed lines for reference. 
Not surprisingly, these angles depend on $\sqrt{s}$, so if we want to make a 
prediction for the CKM angles, we have to further specify the energy scale.
For concreteness, we fix $\sqrt{s}$ at the scale that yields the smallest value of $\mathcal{E}^\perp_{\rm min}$, restricted to the range  $m_Z \lesssim \sqrt{s} \lesssim m_t$.
This occurs at $\sqrt{s} \simeq m_t$, where the top quark cannot participate in the scattering.
Therefore, we find roughly the same minimum as the two-generation study in \Sec{sec:2FS}:
\begin{equation} \label{eq:CKM_bfp1}
    \theta_{\rm{CKM},12}^{\rm min} \sim 6^{\circ}, \quad \theta_{\rm{CKM},23}^{\rm min} \sim \theta_{\rm{CKM},13}^{\rm min} \approx 0\, .
\end{equation}
Since $\theta_{\rm{CKM},13} \approx 0$ (at least for this leading order analysis), this minimization is independent of the complex phase $\delta_{\rm CKM}$, which appears multiplied by $\sin\theta_{\rm{CKM},13}$ in the standard CKM parametrization of \Eq{eq:SP}.
These predicted CKM values can be compared to the measured angles~\cite{ParticleDataGroup:2022pth}:
\begin{equation}
\theta_{\rm{CKM},12}^{\rm exp} \sim 13^{\circ}, \quad \theta_{\rm{CKM},23}^{\rm exp} \sim 2^{\circ}, \quad \theta_{\rm{CKM},13}^{\rm exp} \sim 0.2^{\circ} \, .
\end{equation}

Since we do not have a principle to set the scattering energy, it is interesting to study the entangling behavior as a function of $\sqrt{s}$.
At energies far below the electroweak scale, the flavor-preserving photon exchange dominates, so entanglement suppression no longer enforces a flavor-diagonal coupling of the charged current, leading to maximal mixing between the first two generations.
Above but close to the top quark mass, $\sqrt{s} \gtrsim m_t$, the mass differences between the fermions become negligible and we get roughly equal mixing angles around $4^{\circ}$.
In the high-energy limit with $\sqrt{s} \gg m_t$, the $u \bar{d} \to u \bar{d}$ channel becomes the one that minimizes the entangling power, which prefers no quark mixing.

\subsection{Towards the Full PMNS Matrix}
\label{sec:3FS_PMNS}

\begin{figure*}
\centering
\subfloat[]{
    \includegraphics[width=0.45\textwidth]{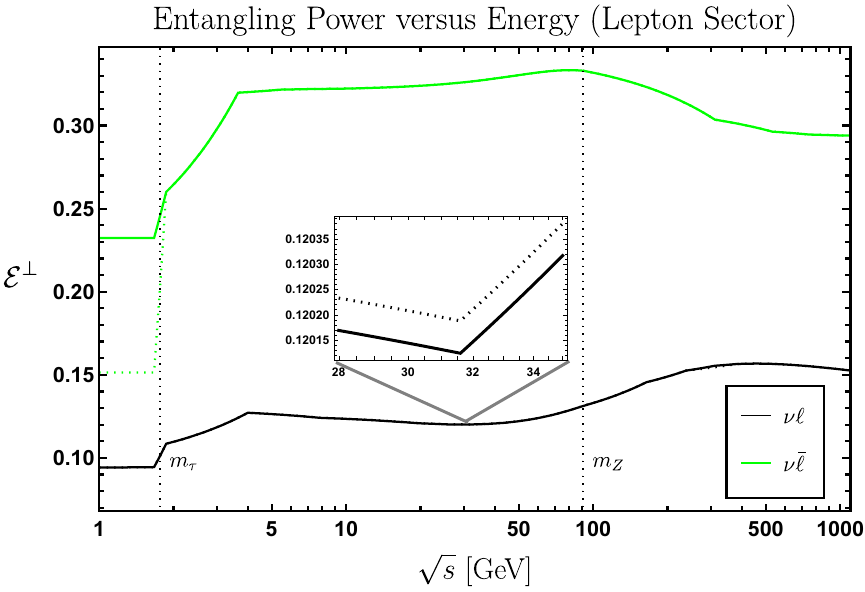} 
    \label{fig:EP_vs_energy_leptons}
}
$\qquad$
\subfloat[]{
 \includegraphics[width=0.45\textwidth]{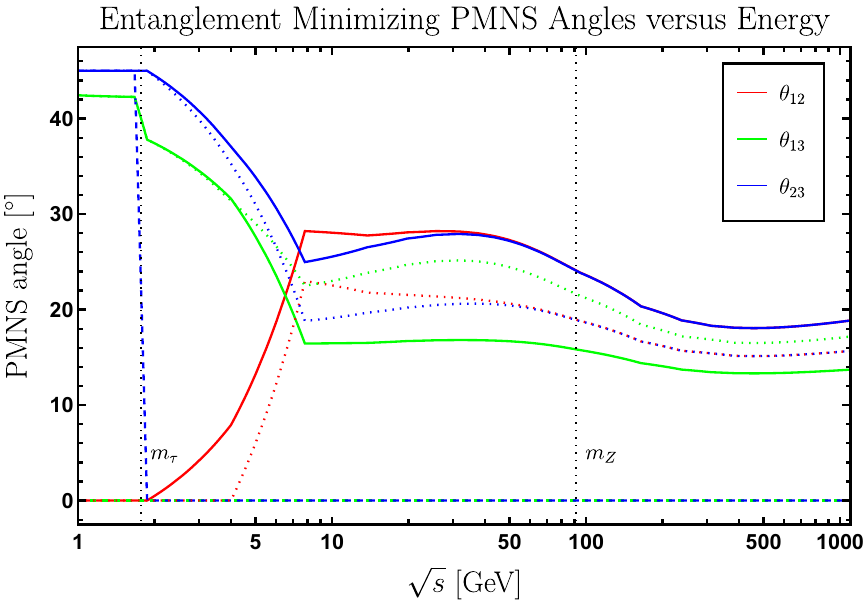}
    \label{fig:angles_vs_energy_leptons}
}
\caption{
Electroweak entanglement in the lepton sector as a function of the scattering center-of-mass energy $\sqrt{s}$.
(a)
The perpendicular entangling power after minimizing over the PMNS angles, considering the two scattering channels $\nu\ell \to \nu\ell$ and $\nu\bar{\ell} \to \nu\bar{\ell}$ separately.
The inset focuses on the region around the minimizing energy to illustrate the effect of the CP-violating phase for the benchmarks $\delta_{\rm{PMNS}}^{\rm{NO}} = 197^{\circ}$ (solid line) and $\delta_{\rm{PMNS}}^{\rm{IO}} = 286^{\circ}$ (dotted line).
(b)
The PMNS angles at the entangling power minimum, with solid lines corresponding to the $\nu\ell \to \nu\ell$ channel that minimizes $\mathcal{E}^{\perp}$ after fixing the CP-violating phase at $\delta_{\rm{PMNS}}^{\rm{NO}} = 197^{\circ}$, dotted lines to the same but at $\delta_{\rm{PMNS}}^{\rm{IO}} = 286^{\circ}$, and dashed lines to minimizing the $\nu\bar{\ell} \to \nu\bar{\ell}$ channel  for reference.
}
\end{figure*}

The analysis for the lepton case mirrors that of the quark case.
In \Fig{fig:EP_vs_energy_leptons}, we show the entangling power of the  $\nu\ell \to \nu\ell$ and $\nu\bar{\ell} \to \nu\bar{\ell}$ channels after separate minimization over the PMNS angles.
Across the whole $\sqrt{s}$ range, the $\nu\ell \to \nu \ell$ channel consistently yields the minimum entangling power.

In contrast to the CKM case, the complex phase $\delta_{\rm{PMNS}}$ affects the value and position of the minimum.
To gain an intuition for its impact, we scan $\delta_{\rm{PMNS}}$ in the range $[0,2\pi]$ at each $\sqrt{s}$ and determine the minimum of $\mathcal{E}_{\nu\ell}^{\perp}$ over the other angles.
In general, we find the rough dependence $\mathcal{E}_{\nu\ell}^{\perp} \sim \sin^2 \delta_{\rm{PMNS}}$ across all energies, which favors minimal CP violation at $\delta_{\rm{PMNS}} \simeq 0 \text{ or } \pi$.
One could view the suppression of CP violation in the neutrino sector as a ``prediction'' of this framework, assuming that higher-order effects do not change the location of the minimum.

There is an intriguing interplay between $\delta_{\rm{PMNS}}$ and the neutrino mass hierarchy.
Global analyses~\cite{Esteban:2020cvm} provide two different best-fit PMNS points depending on normal ordering (NO, $m_{\nu,1}<m_{\nu,2}<m_{\nu,3}$) versus inverted ordering (IO, $m_{\nu,3}<m_{\nu,1}<m_{\nu,2}$) of the neutrino masses.
The NO best fit of $\delta_{\rm{PMNS}}^{\rm{exp, NO}} = {197^{\circ}}^{+41^{\circ}}_{-25^{\circ}}$ is consistent with the entangling minimum at $\pi$, while the IO best fit of $\delta_{\rm{PMNS}}^{\rm{exp, IO}} = {286^{\circ}}^{+27^{\circ}}_{-32^{\circ}}$ is close to the maximum at $3\pi/2$.
As shown with the dotted lines in \Fig{fig:EP_vs_energy_leptons}, the difference in the value of $\mathcal{E}_{\nu\ell}^{\perp}$ between the minimum (solid black) and maximum (dotted black) is relatively small and could be modified by higher-order effects.
Taken seriously, though, this analysis would further ``predict'' normal ordering of neutrinos with minimal CP violation.%
\footnote{Note that \Reff{Quinta:2022sgq} also comes to the same conclusion from an entanglement minimization principle applied to neutrino oscillations.
Given the different origin of entanglement generation from the one considered in this work, we do not discern at present any underlying connection between these two observations.}

The values of the PMNS angles that minimize the entangling power are shown in \Fig{fig:angles_vs_energy_leptons}.
For most of the range $m_{\tau} \lesssim \sqrt{s} \lesssim m_Z$, the minimum happens when all of the angles are sizeable, in agreement with the qualitative experimental picture of lepton mixing.
The energy scale that minimizes $\mathcal{E}_{\nu\ell}^{\perp}$ in this range is $\sqrt{s} \approx 30~\rm GeV$.
Fixing this scale and setting $\delta_{\rm PMNS}$ to the NO best fit value, we find the following PMNS angles:
\begin{equation} \label{eq:PMNS_NO_bfp}
    \theta_{\rm{PMNS},12}^{\rm NO,min} \sim \theta_{\rm{PMNS},23}^{\rm NO,min} = 29^{\circ}\, , \quad  ~\theta_{\rm{PMNS},13}^{\rm NO,min} = 16^{\circ}\, .
\end{equation}
Taking $\delta_{\rm PMNS}$ to be the IO best fit value, we again find that two of the angles are similar, but with a reverse ordering to the third:
\begin{equation} \label{eq:PMNS_IO_bfp}
    \theta_{\rm{PMNS},12}^{\rm IO,min}\sim \theta_{\rm{PMNS},23}^{\rm IO,min} = 21^{\circ}\,, \quad  ~\theta_{\rm{PMNS},13}^{\rm IO,min} = 25^{\circ}\, .
\end{equation}
These PMNS predictions can be compared to the experimental values:
\begin{equation}
\theta_{\rm{PMNS},12}^{\rm exp}\sim 33^{\circ}, 
\quad \theta_{\rm{PMNS},23}^{\rm exp}\sim 48^{\circ}, 
\quad \theta_{\rm{PMNS},13}^{\rm exp}\sim 8^{\circ} \, .
\end{equation}

Compared to the quark case, it is somewhat easier to understand the qualitative behavior as a function of $\sqrt{s}$.
Because the photon channel does not contribute, the only competition is between diagrams with $W$ or $Z$ exchange, which is present at all energy scales.
Balancing these diagrams happens when mixing angles are of comparable size.
Below the tau threshold $\sqrt{s} \lesssim m_\tau$, the three neutrinos scatter against two charged leptons, and balancing flavor-preserving against flavor-violating scattering requires two nearly maximal angles.
Far enough above the tau threshold, $\sqrt{s} \gg m_\tau$, all of the mixing angles need to be around the same size to minimize the entangling power.

\section{Conclusions}
\label{sec:conclusions}

In this paper, we explored the implications of entanglement suppression in electroweak scattering on flavor physics.
By minimizing the perpendicular entangling power, we found a surprising connection between two \emph{a priori} unrelated phenomena in the SM: mixing of the $SU(2)_L$ and $U(1)_Y$ gauge fields and mixing of fermions in flavor space.
Fundamentally, this connection arises because the Weinberg angle encodes the difference in strength between flavor-changing and flavor-preserving processes, which in turn impacts the degree of entanglement.
Because quarks and leptons have different electroweak changes, minimizing entanglement yields different predictions for the respective CKM and PMNS matrices.
Remarkably, this approach yields a CKM matrix that is nearly diagonal but with a non-trivial Cabbibo angle and a PMNS matrix that exhibits large mixing angles, in qualitative agreement with observation.

Given the numerous choices that we made in our definition of entangling power, it is unclear how seriously to take our quantitative ``predictions.''
It is certainly intriguing that the resulting Cabibbo angle of $6^{\circ}$ is only a factor of 2 away from the SM value of $13^{\circ}$, especially because most flavor models can only constrain mixing angles up to $\mathcal{O}(1)$ factors.
It is similarly intriguing that our framework seems to prefer the absence of CP violation in the neutrino sector, though the impact of $\delta_{\rm PMNS}$ is more subtle than the other neutrino mixing angles.
Of course, if this framework is truly enough to explain also quantitatively the experimentally determined values of the flavor-mixing parameters, then one would expect that some kind of higher-order effects would induce the necessary modification of the leading-order result.

It is worth emphasizing the conceptual difference between our analysis and traditional flavor models.
Most flavor models involve new high-energy degrees of freedom, which often lie beyond the reach of the current generation of particle physics experiments.
Here, we did not invoke any new physics degrees of freedom, but nevertheless found relationships between the 19+7 free parameters of the SM plus its neutrino extension.
Ordinarily, flavor models have ``conservation of free parameters,'' such that each adjustable parameter in the SM has (at least) one counterpart in the high-energy description.
By contrast, our approach is based on minimization, which can yield simultaneous predictions for 3 quark and 3 lepton mixing angles (and possibly 1 CP-violating phase in the lepton sector).

There are many potential ways to extend our analysis.
We focused on electroweak scattering in the broken phase, leaving the fermion masses fixed.
In principle, one should consider the entirety of the fermion Yukawa matrices in the minimization process, since these are the quantities that appear in the unbroken phase of the Lagrangian.
The fixed mass approximation is justified a posteriori in our setup, since interesting patterns emerge at energy scales where the external masses are (almost) negligible.
In other words, as long as $\sqrt{s}$ is large enough, including the eigenvalues of the Yukawa matrices in the minimization process would not have any impact on our conclusions.
Allowing the masses to vary in general could alter the picture close to the various mass thresholds, however, and it would be interesting to study this in more detail in future work.
One would hope that the choices we had to make in this paper could be derived from a overarching principle, and perhaps spontaneous symmetry breaking is somehow related to the scale choice when considering the interplay between fermion masses and mixings.

Ultimately, we want to know:  \emph{is this all just a numerical coincidence, or could minimization of quantum entanglement really be a fundamental principle of nature?}
To answer this question, we encourage further explorations of the consequences of entanglement in various scattering processes, both in and beyond the SM.
While previous studies have focused on the emergence of symmetries, our analysis shows that entanglement suppression can lead to predictions for symmetry-breaking effects, suggesting that other kinds of consequences could arise from this simple principle.
Though speculative, this line of research highlights the potential of injecting concepts from quantum information theory into high-energy physics.

\begin{acknowledgments}

We thank Gilly Elor, Darius Faroughy, Yuval Grossman, Gino Isidori, Ian Low, Grant Remmen, Olcyr Sumensari, Raymond Volkas, and Carlos Wagner for useful discussions and appropriately skeptical feedback on our work.
This work was supported by the Office of High Energy Physics of the U.S. Department of Energy (DOE) under Grant No.~DE-SC0012567, and by the DOE QuantISED program through the theory consortium “Intersections of QIS and Theoretical Particle Physics” at Fermilab (FNAL 20-17).
J.T.\ is additionally supported by the DOE Office of Science, National Quantum Information Science Research Centers, Co-design Center for Quantum Advantage (C$^2$QA) under Contract No.~DE-SC0012704, by the Simons Foundation through
Investigator grant 929241, and his work was performed in part at the Aspen Center for Physics, which is supported by the National Science Foundation grant PHY-2210452.
S.T. was additionally supported by the Swiss National Science Foundation - project n.~P500PT\_203156.

\end{acknowledgments}

\begin{appendix}

\section{Description of Scattering/Measurement Process}
\label{app:scattering_details}

In this appendix, we present a complete derivation of the final-state density matrix used in our calculations.
In general, scattering processes are described by means of the $\mathcal S$ operator, which connects the Fock spaces $\mathcal F$ of all incoming and outgoing asymptotic states:
\begin{equation} \label{eq:out=Sin}
   \ket{\rm{out}} = \mathcal S \ket{\rm{in}}\,.
\end{equation}
The $\mathcal S$ operator is unitary and it can be defined in terms of the transfer operator $\mathcal T$:
\begin{equation} \label{eq:S-matrix_def}
    \mathcal{S} \equiv \mathbb{I} + i \mathcal{T}\,.
\end{equation}

For our purposes, we care about matrix elements between initial and final states in $2 \to 2$ processes, which are encoded in the S-matrix:
\begin{equation}
   \label{eq:S-matrix}
   S_{k\ell ij} =  \bra{p_3,k;p_4,\ell}\mathcal{S}\ket{p_1,i;p_2,j}\, . 
\end{equation}
The inner product is defined as
\begin{align}
   \label{eq:inner_product}
   &\braket{p_3,k;p_4,\ell | p_1,i;p_2,j} \notag \\
   &= 2E_3 2E_4 (2\pi)^3 \delta^{(3)}(\vec{p}_3-\vec{p}_1) (2\pi)^3 \delta^{(3)}(\vec{p}_4-\vec{p}_2)\delta_{ik}\delta_{j\ell}\, ,
\end{align}
and the perturbative amplitudes $\mathcal M$ can be extracted from the transfer operator:
\begin{align}
   \label{eq:T-matrix}
   &\bra{p_3,k;p_4,\ell}i \, \mathcal T \ket{p_1,i;p_2,j} \notag \\
   &= i (2\pi)^4\delta^{(4)}(p_1+p_2-p_3-p_4) \mathcal M_{k\ell ij}(p_1,p_2\to p_3,p_4)\, ,
\end{align}
where we work at tree-level for our analysis.

For the discussion below, we focus on the $ud \to ud$ channel, which has an obvious generalization to $u\bar{d} \to u \bar{d}$.
Following \Eq{eq:flavor_kinematics}, quark flavors and momenta are labeled as: 
\begin{equation} 
u_{Li}(p_1) d_{Lj}(p_2) \to u_{Lk}(p_3) d_{L\ell}(p_4)\, .
\end{equation}
For convenience, we work in the center-of-mass frame with $\vec{p}_1 = - \vec{p}_2$ and $p_1+p_2=(\sqrt{s},0,0,0)^{\rm{T}}$.

As discussed in \Sec{sec:entangling_power}, we focus on elastic scattering within the finite-dimensional Hilbert space $H_f$.
To map from the full Fock space $\mathcal{F}$ to the flavor Hilbert space $H_f$, we define two projection operators:
\begin{enumerate}
    \item[i)] $\Pi_{\rm{in}}$ prepares an unentangled initial state of two particles with definite momentum and helicity:
\begin{align} \label{eq:in}
    \ket{\rm{in}}_{ij} &= \ket{p_1,i; p_2,j} \notag \\
    &= \sqrt{2 E_1} (a_{\vec{p}_1,i}^L)^{\dagger} \ket{0}_u \otimes \sqrt{2 E_2} (b_{\vec{p}_2,j}^L)^{\dagger} \ket{0}_d\, ,
\end{align}
where the creation and annihilation operators obey the usual anticommutation relations. The label $L$ denotes eigenstates of negative helicity.
    \item[ii)] $\Pi_{\rm{out}}$ corresponds to the final state measurement.
    First, we restrict the final state to have two particles \cite{Peschanski:2016hgk} with negative helicity, a combination of actions denoted by $\Pi_2^L$.
    Then, we measure the scattering angle $\Theta$ of one of the final state particles, which without loss of generality we choose to be the up-type quark.%
\footnote{Even though the scattering amplitude is independent of the azimuthal angle $\Phi$, an actual measurement of the final state particle in real space would involve the placement of a detector at a definite $\Phi$ value. We assume that such an angle is arbitrarily selected in our idealized measurement procedure.}
    The full measurement can be written schematically as:
\begin{equation} \label{eq:Pi_out}
\Pi_{\rm{out}} \equiv (\ket{\Theta}\bra{\Theta}_u \otimes \mathbb{I}_d) \circ \Pi_2^L\, ,
\end{equation}
such that the post-measurement state is:
\begin{equation} \label{eq:out}
    \ket{\rm{out}}_{ij} = \frac{\Pi_{\rm{out}} \mathcal S \ket{\rm{in}}_{ij}}{\left|\Pi_{\rm{out}} \mathcal S \ket{\rm{in}}_{ij}\right|} \,.
\end{equation}
Crucially, this projection does not involve any measurement of flavor.

\end{enumerate}

Performing the above projections, the post-measurement state can be written as:
\begin{align} \label{eq:theta_out}
   \ket{\rm out}_{ij} = \frac{1}{\mathcal N_{ij}} \sum_{k,\ell=1}^G \mathcal{M}_{k\ell ij}\left(s,\Theta\right) \ket{p_3,k;p_4,\ell}\, , 
\end{align}
Here, we have made explicit that the scattering amplitude is a function of the center-of-mass energy $\sqrt{s}$ and scattering angle $\Theta$, and we assume that $\Theta \not=0$ to avoid forward scattering contributions.
The normalization factor $\mathcal N_{ij}$ is defined such that $|\ket{\rm out}_{ij}| = 1$.

Having fixed the kinematics, the final state depends only on our choice for the initial flavor eigenstates, namely $\ket{ij}_{ud} \in H_f$.
We can thus identify the matrix elements of $\mathcal{S}_f$ from \Sec{sec:entangling_power}:
\begin{equation}\label{eq:Sf_def}
\bra{k \ell}_{ud} \mathcal{S}_f \ket{ij}_{ud} = \frac{\mathcal{M}_{k\ell ij}(s,\Theta)}{\mathcal{N}_{ij}} \,
\end{equation}
This flavor scattering operator is a $G^2 \times G^2$-dimensional matrix acting on the product states in \Eq{eq:generic_state}.
Setting $\Theta = \pi/2$, we obtain the operator $\mathcal{S}_f^\perp$ appearing in \Eq{eq:perp_entangle_power}.

Schematically, the various scattering and projection operators act as follows:
\begin{equation}
\begin{array}{ccc}
 \mathcal F & \xrightarrow{\hspace{0.2cm}\mathcal{S}\hspace{0.2cm}} & \mathcal F \\
\quad \Big\downarrow  \Pi_{\rm in}  &  & \quad \Big\downarrow \Pi_{\rm out} \\
H_{f} & \xrightarrow{\hspace{0.2cm}\mathcal{S}_f\hspace{0.2cm}} & H_{f}
\end{array}\,.
\end{equation}
While $\mathcal{S}_f$ is not a unitary operator, it still preserves the normalization of initial states with definite flavor, i.e.\ $\text{diag}\big(\mathcal{S}_f \mathcal{S}_f^\dagger \big)= \mathbb{I}$.

To compute the linear entropy in \Eq{eq:lin_entropy}, we need the reduced density matrix in flavor space.
Note that there is a separate density matrix for each initial state $\ket{ij}_{ud}$.
The density matrix associated with \Eq{eq:theta_out} is:
\begin{equation}
    \rho_{ij}  \equiv \big(\mathcal{S}_f \ket{ij}_{ud}\big) \big(\bra{ij}_{ud} \mathcal{S}_f^\dagger\big)\,,
\end{equation}
where there is \emph{not} a sum over $ij$.
By tracing out the down-quark degrees of freedom, the matrix elements of the reduced density matrix $\rho_{R,ij} \equiv \text{Tr}_{u} \rho_{ij}$ are:
\begin{align} \label{eq:red_density}
&\bra{k}_u \rho_{R,ij} \ket{k'}_u =\frac{1}{|\mathcal N_{ij}|^2} \sum_{\ell=1}^G \mathcal{M}_{k\ell ij}(s,\Theta) \, \mathcal{M}^*_{k'\ell ij}(s,\Theta) \, . 
\end{align}

\section{Parametrization of CKM and PMNS}
\label{app:standard_parametrization}

The standard parametrization of the CKM and PMNS matrices is:
\begin{align} \label{eq:SP}
  &V = \notag \\
  &\begin{pmatrix}
c_{12} c_{13} & s_{12} c_{13} & s_{13} e^{-i \delta} \\
-s_{12} c_{23} - c_{12} s_{23} s_{13} e^{i \delta} & c_{12} c_{23} - s_{12} s_{23} s_{13} e^{i \delta} & s_{23} c_{13} \\
s_{12} s_{23} - c_{12} c_{23} s_{13} e^{i \delta} & -c_{12} s_{23} - s_{12} c_{23} s_{13} e^{i \delta} & c_{23} c_{13}
\end{pmatrix}\, 
\end{align}
where $c_{ij}=\cos\theta_{ij}$ and $s_{ij}=\sin\theta_{ij}$.
The mixing angles are restricted to the range $\theta_{ij} \in [0,\pi/2]$, and the CP-violating phase satisfies $\delta \in [0,2\pi]$.

According to global fits, the CKM angles are~\cite{ParticleDataGroup:2022pth}:
\begin{align} \label{eq:CKM_bfp2}
    \theta_{\rm{CKM},12}^{\rm exp} &= 13.002^{\circ}\pm 0.038^{\circ}\, ,\notag\\
    \theta_{\rm{CKM},23}^{\rm exp} &= 2.396^{\circ}\pm 0.048^{\circ}\, ,\notag\\
    \theta_{\rm{CKM},13}^{\rm exp} &= 0.211^{\circ}\pm 0.006^{\circ}\, ,\notag\\
    \delta^{\rm exp}_{\rm{CKM}} &= 65.546^{\circ}\pm 1.546^{\circ}\, .
\end{align}
Similarly, the global fits for the PMNS angles are~\cite{Esteban:2020cvm}:
\begin{align} \label{eq:PMNS_bfp}
    \theta_{\rm{PMNS},12}^{\rm exp} &= 33.82^{\circ}\pm 0.78^{\circ}\, , \notag \\
    \theta_{\rm{PMNS},23}^{\rm exp} &= 48.3^{\circ}\pm 1.4^{\circ}\, , \notag \\
    \theta_{\rm{PMNS},13}^{\rm exp} &= 8.61^{\circ}\pm 0.13^{\circ}\, , \notag \\
    \delta_{\rm{PMNS}}^{\rm{exp, NO}} &= {197^{\circ}}^{+41^{\circ}}_{-25^{\circ}}\, , \notag \\
    \delta_{\rm{PMNS}}^{\rm{exp, IO}} &= {286^{\circ}}^{+27^{\circ}}_{-32^{\circ}}\, ,
\end{align}
with different preferred ranges for $\delta_{\rm{PMNS}}$ depending on whether neutrino masses exhibit normal ordering (NO) or inverted ordering (IO).

\section{Electroweak Scattering Amplitudes}
\label{app:amplitudes}

We now calculate the scattering amplitude $\mathcal M$ for the electroweak processes analyzed in the main text.

Starting with the kinematics, we parametrize the three-momenta of the external particles in spherical coordinates as: 
\begin{equation}
\vec{p} = \left| \vec{p} \right| (\sin\theta \cos\phi, \sin\theta \sin\phi, \cos\theta)^{\rm{T}},
\end{equation}
and use the following conventions in the center-of-mass frame:
\begin{align} \label{eq:com_frame}
p_1:& \quad \theta_1=0, \quad \phi_1=0\, , \notag\\
p_2:& \quad \theta_2=\pi, \quad \phi_2=\pi\,, \notag\\
p_3:& \quad \theta_3=\Theta, \quad \phi_3=0\, ,\notag\\
p_4:& \quad \theta_4=\pi-\Theta, \quad \phi_4=\pi\, .
\end{align}
We only consider incoming and outgoing fermionic states that are described by negative helicity spinors $u_{L}$ for particles and positive helicity spinors $v_{R}$ for anti-particles. 
In the Dirac representation, the four-component spinors in the helicity basis are:\footnote{The spinors are eigenstates of the helicity operator
\[
\hat\lambda = \frac{\vec{\Sigma} \cdot \vec{p}}{|\vec{p}|}\,, \qquad \Sigma_i = \frac{1}{2} \begin{pmatrix}
\sigma_i & 0 \\
0 & \sigma_i 
\end{pmatrix}\, ,
\]
where $\vec{\Sigma}$ is the spin operator and $\sigma_i$ are the Pauli matrices.}
\begin{align} \label{eq:spinors}
&u_{L}(p) = \sqrt{E + m} \begin{pmatrix} 
-\sin\frac{\theta}{2} \\ 
e^{i\phi} \cos\frac{\theta}{2}  \\
\frac{|\vec{p}|}{E + m} \sin\frac{\theta}{2} \\
-\frac{|\vec{p}|}{E + m} e^{i\phi} \cos \frac{\theta}{2} 
\end{pmatrix}\, , \notag \\
&v_{R}(p)  = \sqrt{E + m} \begin{pmatrix} 
\frac{|\vec{p}|}{E + m} \sin \frac{\theta}{2} \\
-\frac{|\vec{p}|}{E + m} e^{i\phi} \cos \frac{\theta}{2}  \\
-\sin \frac{\theta}{2} \\
e^{i\phi} \cos \frac{\theta}{2} 
\end{pmatrix}\, ,
\end{align}
where the angles for each particle are defined according to \Eq{eq:com_frame}.

The amplitudes of the scattering channels in \Fig{fig:diagrams} are
\begin{align} \label{eq:ampl_udud}
&i \mathcal M_{k\ell ij} (u_{L}^i(p_1) d_{L}^j(p_2) \to u_{L}^k(p_3) d_{L}^\ell(p_4)) = \notag \\
&\quad -i \frac{g^2 \sin^2\theta_W Q^uQ^d}{t} \delta_{ik}\delta_{j\ell} \big[\mathcal{J}_{uu}^{\gamma,\mu}\big]_{ki}^{31} \big[\mathcal{J}_{dd,\mu}^{\gamma}\big]_{\ell j}^{42} \notag \\
&\quad -i \frac{g^2}{\cos^2\theta_W }\frac{1}{t-m_Z^2} \delta_{ik}\delta_{j\ell} \big[\mathcal{J}_{uu}^{Z,\mu}\big]_{ki}^{31} \big[\mathcal{J}_{dd,\mu}^{Z}\big]_{\ell j}^{42} \notag \\
&\quad +i \frac{g^2 V_{ij}^{\ast}V_{k\ell}}{2}\frac{1}{u-m_W^2}  \big[\mathcal{J}_{du}^{W,\mu}\big]_{\ell i}^{41} \big[\mathcal{J}_{ud,\mu}^{W}\big]_{k j}^{32}\, ,
\end{align}
where the different signs between the charged- and neutral-currents comes from fermion crossing.
Note that since we are fixing the external fermion masses to their current values and we perform the bulk of our analysis below the top mass, contributions induced by Higgs exchange that are proportional to the small fermion masses are negligible and we, therefore, omit them.

For the process obtained by crossing symmetry
\begin{align} \label{eq:ampl_udbudb}
&i \mathcal M_{k\ell ij} (u_{L}^i(p_1) \bar{d}_{R}^j(p_2) \to u_{L}^k(p_3) \bar{d}_{R}^\ell(p_4)) = \notag \\
&\quad -i \frac{g^2 \sin^2\theta_W Q^uQ^d}{t} \delta_{ik}\delta_{j\ell} \big[\mathcal{J}_{uu}^{\gamma,\mu}\big]_{ki}^{31} \big[\mathcal{J}_{\bar{d}\bar{d},\mu}^{\gamma}\big]_{j \ell}^{24} \notag \\
&\quad -i \frac{g^2}{\cos^2\theta_W }\frac{1}{t-m_Z^2} \delta_{ik}\delta_{j\ell} \big[\mathcal{J}_{uu}^{Z,\mu}\big]_{ki}^{31} \big[\mathcal{J}_{\bar{d}\bar{d},\mu}^{Z}\big]_{j \ell}^{24} \notag \\
&\quad +i \frac{g^2 V_{i\ell}^{\ast}V_{kj}}{2}\frac{1}{s-m_W^2} \big[\mathcal{J}_{\bar{d}u}^{W,\mu}\big]_{j i}^{21} \big[\mathcal{J}_{u\bar{d},\mu}^{W}\big]_{k \ell}^{34}\, .
\end{align}
In these expressions, we defined the bilinears:
\begin{align}
&\big[\mathcal{J}_{qq}^{\gamma,\mu}\big]_{ij}^{IJ}= \bar{u}_L(q^i; p_I)\gamma^\mu  u_L(q^j; p_J) \, , \notag \\ 
&\big[\mathcal{J}_{qq}^{Z,\mu}\big]_{ij}^{IJ}= \bar{u}_L(q^i; p_I)\gamma^\mu (g_V^q-g_A^q\gamma^5)  u_L(q^j; p_J) , \notag \\ 
&\big[\mathcal{J}_{ud}^{W,\mu}\big]_{ij}^{IJ}=  \frac{1}{2}\bar{u}_L(q^i; p_I)\gamma^\mu (1-\gamma^5)  u_L(q^j; p_J)\,,
\end{align}
where $u_L(q^i; p_I)$ denotes the spinor in \Eq{eq:spinors} for the quark $q$ of flavor $i$ and momentum $p_I$.

and for anti-quarks, we exchange the spinors $u_L \leftrightarrow v_R$.
The $Z$ couplings are defined as
\begin{equation} \label{eq:Z_couplings}
   g_V^f = \frac{1}{2}T_3^f - Q^f \sin\theta_W^2\, , \quad g_A^f = \frac{1}{2} T_3^f,
\end{equation}
where the relevant charges for quarks are given in \Eq{eq:quark_charges}.

The Mandelstam variables are given by
\begin{align}  \label{eq:tu_mandelstam}
s &= (E_1+E_2)^2=(E_3+E_4)^2\,, \notag \\
t &= -\frac{s}{2} \Big[1 - \frac{m_1^2 + m_2^2 + m_3^2 + m_4^2}{s} \notag \\
& \qquad\qquad + \frac{(m_1^2 - m_2^2)(m_3^2 - m_4^2)}{s^2} \notag \\
& \qquad\qquad - \lambda\left(\frac{m_1^2}{s}, \frac{m_2^2}{s}\right) \lambda\left(\frac{m_3^2}{s}, \frac{m_4^2}{s}\right) \cos\Theta\Big]\, , \notag \\
u &= -\frac{s}{2} \Big[1 - \frac{m_1^2 + m_2^2 + m_3^2 + m_4^2}{s^2} \notag \\
& \qquad\qquad - \frac{(m_1^2 - m_2^2)(m_3^2 - m_4^2)}{s^2} \notag \\ 
& \qquad\qquad + \lambda\left(\frac{m_1^2}{s}, \frac{m_2^2}{s}\right) \lambda\left(\frac{m_3^2}{s}, \frac{m_4^2}{s}\right) \cos\Theta\Big]\, ,
\end{align}
where $\lambda(x,y) = \sqrt{(1-x-y)^2-4xy}$.

In the limit where all the masses are negligible, $u_{L}$ and $v_{R}$ correspond to left-handed chiral particles and right-handed chiral anti-particles.
The amplitudes in \Eq{eq:ampl_udud} and \Eq{eq:ampl_udbudb} become then pure helicity amplitudes:
\begin{align}
&\big[\mathcal{J}_{uu}^{\gamma,\mu}\big]_{ki}^{31} \big[\mathcal{J}_{dd,\mu}^{\gamma}\big]_{\ell j}^{42} =  \frac{1}{Y^u Y^d} \big[\mathcal{J}_{uu}^{Z,\mu}\big]_{ki}^{31} \big[\mathcal{J}_{dd,\mu}^{Z}\big]_{\ell j}^{42} \notag \\
&\quad =-\big[\mathcal{J}_{du}^{W,\mu}\big]_{\ell i}^{41} \big[\mathcal{J}_{ud,\mu}^{W}\big]_{k j}^{32}  = 2s\,, \notag \\
&\big[\mathcal{J}_{uu}^{\gamma,\mu}\big]_{ki}^{31} \big[\mathcal{J}_{\bar{d}\bar{d},\mu}^{\gamma}\big]_{j \ell}^{24} =  \frac{1}{Y^u Y^d} \big[\mathcal{J}_{uu}^{Z,\mu}\big]_{ki}^{31} \big[\mathcal{J}_{\bar{d}\bar{d},\mu}^{Z}\big]_{j \ell}^{24} \notag \\
&\quad = - \big[\mathcal{J}_{\bar{d}u}^{W,\mu}\big]_{j i}^{21} \big[\mathcal{J}_{u\bar{d},\mu}^{W}\big]_{k \ell}^{34} = 2u\,,
\end{align}
where we have used the Fierz identities in the third equalities of each equation.

Throughout the analysis, we employ for the QED and EW couplings the corresponding one-loop RG evolution.
The numerical boundary conditions are taken at $\sqrt{s} = m_Z = 91.187~\rm{GeV}$ to be $\alpha_1=0.0101$ and $\alpha_2 = 0.034$.

Finally, one can obtain the amplitudes of the lepton-neutrino scattering by replacing $u \leftrightarrow \nu$ and $d \leftrightarrow \ell$ and the respective charges with
\begin{align} \label{eq:lepton_charges}
Q^\ell = -1\, , ~ Q^\nu = 0\, , ~ Y^\ell = -\frac{1}{2} + \sin\theta_W^2\, , ~ Y^\nu = \frac{1}{2}\, .
\end{align}

\end{appendix}

\bibliographystyle{JHEP}
\bibliography{main}
\end{document}